\documentclass{mnras}

\usepackage{rotating}
\usepackage{hyperref}
\usepackage{multirow}

\usepackage{apalike}

\usepackage{natbib}

\begin{document}

\title{Quasi-Periodic Oscillations in the long term radio light curves of the blazar AO 0235+164}

\author[Tripathi et al.]{Ashutosh Tripathi$^{1}$, Alok C.\ Gupta$^{2,3}$, Margo F. Aller$^{4}$, Paul J.\ Wiita$^{5}$,
\newauthor
Cosimo Bambi$^{1}$, Hugh Aller$^{4}$, Minfeng Gu$^{2}$\\
\\
$^{1}$Center for Field Theory and Particle Physics and Department of Physics, Fudan University, 2005 Songhu Road, 200438 Shanghai, \\
China\\
$^{2}$Key Laboratory for Research in Galaxies and Cosmology, Shanghai Astronomical Observatory, Chinese Academy of Sciences, \\
80 Nandan Road, Shanghai 200030, China \\
$^{3}$Aryabhatta Research Institute of Observational Sciences (ARIES), Manora Peak, Nainital -- 263001, India\\
$^{4}$Department of Astronomy, University of Michigan, 311 West Hall, Ann Arbor, MI 48109-1107, USA\\
$^{5}$Department of Physics, The College of New Jersey, P.O.\ Box 7718, Ewing, NJ 08628-0718, USA
}

\maketitle
\begin{abstract}

We present time series analyses of three-decade long radio observations of the BL Lacertae object AO 0235+164 made at the University of Michigan Radio Astronomical Observatory operating at three central frequencies of 4.8 GHz, 8.0 GHz and 14.5 GHz. We detected a quasi-periodic oscillation of $\sim$965 days in all three frequency bands in the light curve of the effectively simultaneous observations, along with strong signals at $\sim$1950 d, $\sim$1350 d, and $\sim$660 d. The periodicity is analyzed with three methods: Data Compensated Discrete Fourier Transform, Generalized Lomb-Scargle Periodogram and Weighted Wavelet Z-transform. These methods are chosen as they have different analysis approaches toward robust measurement of claimed periodicities. The QPO at $965\pm 50$ days is found to be significant (at least $3.5\sigma$) and is persistent throughout the observation for all three radio frequencies, and the others, which may be harmonics, are comparably significant in at least the 8.0 GHz and 14.5 GHz bands. We briefly discuss plausible explanations for the origin of such long and persistent periodicity. 
\end{abstract}


\section{Introduction}

\noindent
Quasi-periodic oscillations (QPOs) are fairly common in the X-ray time series data of stellar mass X-ray black 
hole (BH) and neutron star binaries in the Milky Way and nearby galaxies \citep[e.g.,][]{2006ARA&A..44...49R} but 
have  been rarely detected in various subclasses of active galactic nuclei (AGN) 
\citep[e.g.,][and references therein]{2014JApA...35..307G,2018Galax...6....1G,2019MNRAS.484.5785G}. The 
centers of AGNs host accreting supermassive black holes (SMBHs) in the mass range 10$^{6}$ -- 10$^{10}$ M$_{\odot}$ 
and have several similarities with the scaled up stellar mass X-ray BH and neutron star binaries in the 
Milky Way and nearby galaxies. So, searches for QPOs in the time series data of different electromagnetic 
(EM) bands of AGNs are  important to better understand emission models of AGNs. \\ 
\\
The blazar subclass of radio-loud AGNs consists of BL Lacertae (BL Lac) objects and flat spectrum radio 
quasars (FSRQs). Blazars show large flux and spectral variability in the entire electromagnetic (EM) spectrum,  ranging 
from radio to GeV and sometimes even TeV $\gamma-$rays.  Significant and variable polarization is seen in radio and optical bands. Blazars 
have compact radio cores and their radiation across all EM bands is predominantly non-thermal.  Blazar properties can be fundamentally understood
if they possess relativistic charged particle jets that make an angle of $<10^{\circ}$ from the observer's line of
sight \citep{1995PASP..107..803U}.  Blazar Spectral Energy Distributions (SEDs) have double 
humped structures in which the low energy hump peaks lies between IR and X-rays whereas the high energy hump extends over the 
$\gamma-$ray bands. The low energy SED hump is dominated by synchrotron radiation while the high energy one is usually attributed to
inverse Compton (IC) radiation although hadronic models are often also possible \citep[e.g.,][]{2007Ap&SS.309...95B}. \\      
\\
There have been  claims of detection of periodic or quasi-periodic variation in time series data of AGN  reported from different EM bands on diverse
time scales \citep[see for reviews,][and references therein]{2014JApA...35..307G,2018Galax...6....1G}. 
Thanks to the newer techniques for searching for periodic oscillations and/or QPOs with significance tests, 
since 2008 there have been  somewhat stronger claims of such detections in blazars 
and other classes of AGN  \citep[e.g.,][and references therein]{2008Natur.455..369G,2009A&A...506L..17L,2009ApJ...690..216G,2018A&A...616L...6G,2019MNRAS.484.5785G,2013MNRAS.436L.114K,2015ApJ...813L..41A,2017ApJ...847....7B,2018Galax...6..136B}.  \\
\\
AO 0235$+$164 is at redshift $z =$ 0.94 \citep{1996A&A...314..754N} and was one of the first objects classified 
as a BL Lac object \citep{1975ApJ...201..275S}. It  has shown a high degree of polarization in optical/NIR bands 
\citep[$P_V$ = 43.9$\pm$1.4\%, $P_J$ = 39.9$\pm$0.5\%][]{1982MNRAS.198....1I}. With the help of optical photometric and spectroscopic 
observations, it was noticed that this blazar has  absorbing systems from foreground galaxies at $z =$ 0.524 
and $z =$ 0.851 \citep{1976ApJ...205L.117B,1987ApJ...318..577C,1996A&A...314..754N} which makes it a 
gravitationally micro-lensed blazar. It is a bright and highly variable blazar, and its variability has been extensively
studied in many EM bands on  diverse timescales 
\citep[e.g.,][and references therein]{1980ApJ...236...84P,1988ApJ...326..668O,1989A&A...220...65W,1992A&A...266..101S,1995ApJS..100...37G,1996A&A...305...42H,1996ApJ...459..156M,1999ApJS..121..131F,2000A&A...360L..47R,2000ApJ...545..758R,2000AJ....120...41W,2001A&A...377..396R,2003APSC..300..159A,2004A&A...419..913O,2005A&A...438...39R,2006A&A...452..845R,2008A&A...480..339R,2011ApJ...735L..10A,2012ApJ...751..159A,2018MNRAS.475.4994K}.  AO 0235$+$164 exhibits large and variable polarization from radio to optical bands
\citep[e.g.,][and references therein]{1979ApJ...229L...1L,1982MNRAS.198....1I,2011PASJ...63..489S}. Here
we report the detection of a QPO in the gravitationally lensed BL Lac object AO 0235$+$164 with the dominant period
of $\sim$965 days in 4.8 GHz, 8.0 GHz and 14.5 GHz radio observations taken over 32 years made 
at the University of Michigan Radio Astronomy Observatory (UMRAO).  \\
\\
In Section 2, we briefly describe the radio data taken at UMRAO. In Section 3, we 
present the QPO search methods used. The results of those analyses are described in Section 4 
and a discussion and our conclusions are given in Section 5.

\begin{figure*}
\centering
\includegraphics[scale=0.8]{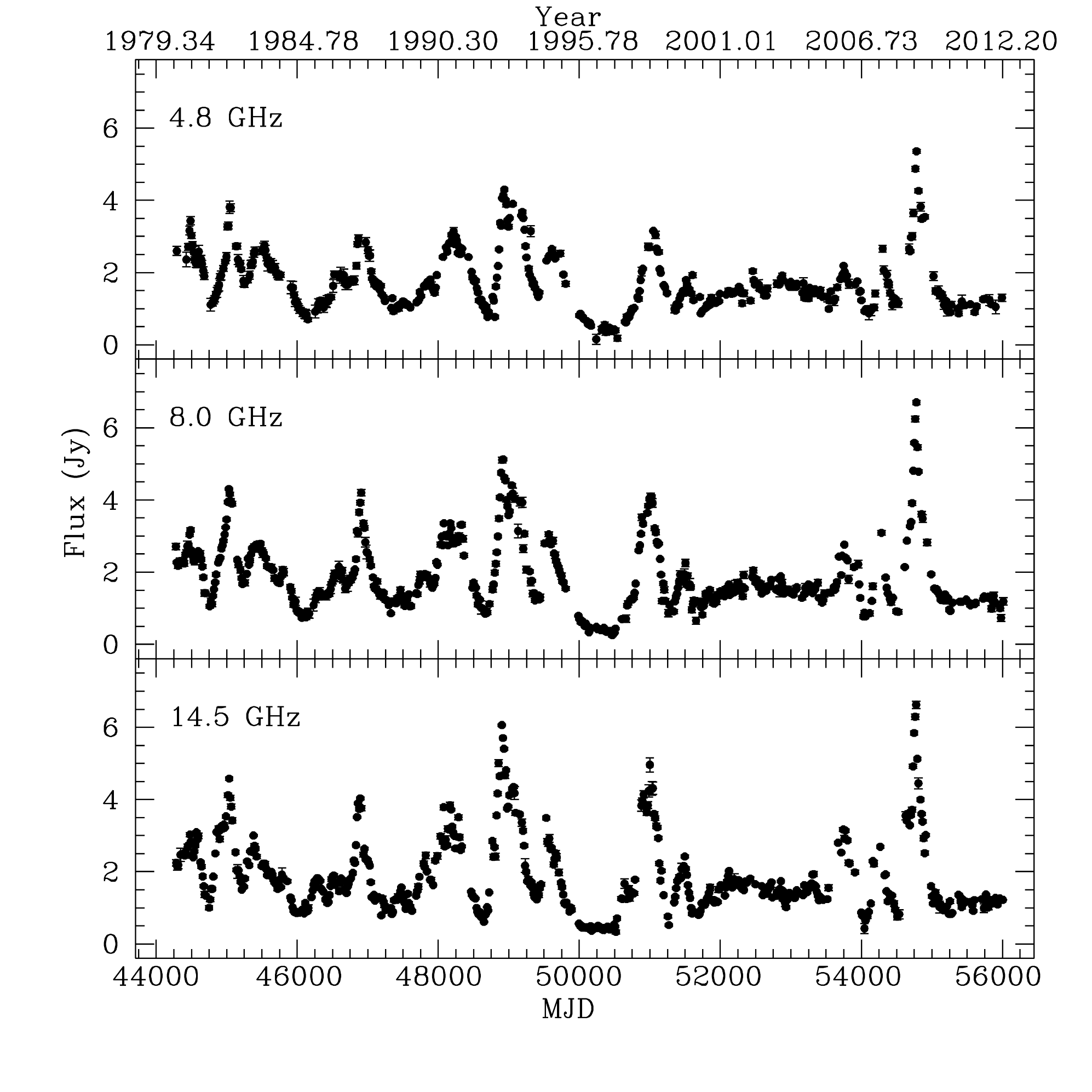}
\caption{UMRAO 15-day binned long-term radio light curves of AO 0235+164 measured between MJD 44270 and MJD {\bf 56012}.}
\end{figure*}

\section{Observations and Data Analysis}

\noindent
The UMRAO total 
flux density observations analyzed were obtained
with the Michigan 26-meter equatorially-mounted, prime focus, paraboloid as
part of the University of  Michigan extragalactic variable source
monitoring program \citep{1985ApJS...59..513A} from January 1980 until closure of this facility in
June 2012. This telescope was equipped with transistor-based radiometers
operating at three primary frequencies centered at 4.8, 8.0, and 14.5 GHz.
At 4.8 GHz a single feed-horn, mode-switching system with a central
operating frequency of 4.80 GHz and a bandwidth of 0.68 GHz was employed. 
At 8.0 GHz, a dual feed horn system was used with an on-on observing technique
and the bandwidth used is 0.79 GHz. The primary frequency of 14.5 GHz employed dual,
linearly polarized, rotating feed horn polarimeters which were placed symmetrically about 
the prime focus of the paraboloid; here, the bandwidth used is 1.68 GHz. \\
\\
The target source was observed regularly as part of this monitoring program
with some gaps due to the annual proximity of the source to the sun or when
extremely poor weather conditions occurred. In general each daily-averaged 
observation of AO 0235+164 consisted of a series of 8 to 16 individual 
measurements obtained over approximately 35 minutes. The adopted flux 
density scale is based on \citet{1977A&A....61...99B} and
used Cassiopeia A (3C 461) as the primary standard.  In addition to the
observations of this primary standard, observations of nearby secondary
flux density calibrators were interleaved with the observations of the
target source every 1.5 to 3 hours to verify the stability of the antenna
gain and to verify the accuracy of the telescope pointing.  Although observations at
14.5 and 8.0 GHz began in 1975, here we consider only the data from 1980 onwards where
data are also available at 4.8 GHz.  

\section{Light Curve Analysis and Results}

\noindent
The three decades long duration radio flux densities of the blazar AO 0235$+$164 at 4.8 GHz, 8.0 GHz, and 14.5 GHz 
are presented in Fig.\ 1. On a visual inspection the light curves show modulations with a possible quasi-periodic component. To estimate the period of the modulation and test its significance, we analysed the light curve using three methods:  
Data Compensated Discrete Fourier Transform (DCDFT); Generalized Lomb-Scargle periodogram (GLSP);	  and 
Weighted Wavelet Z-transform (WWZ). These methods are chosen as they have different approaches to
detecting any periodicities in the light curves and can be used for unevenly sampled data. 
The DCDFT is used for unevenly sampled data to  identify 
 periodicity as peaks  in the power spectral density (PSD). The GLSP is also used for unevenly sampled 
 data but employs sinusoids plus constants as the fitting function.
The WWZ searches for both periodicity in the data and its persistence in time. 

\begin{figure*}
\centering
\includegraphics[scale=0.5]{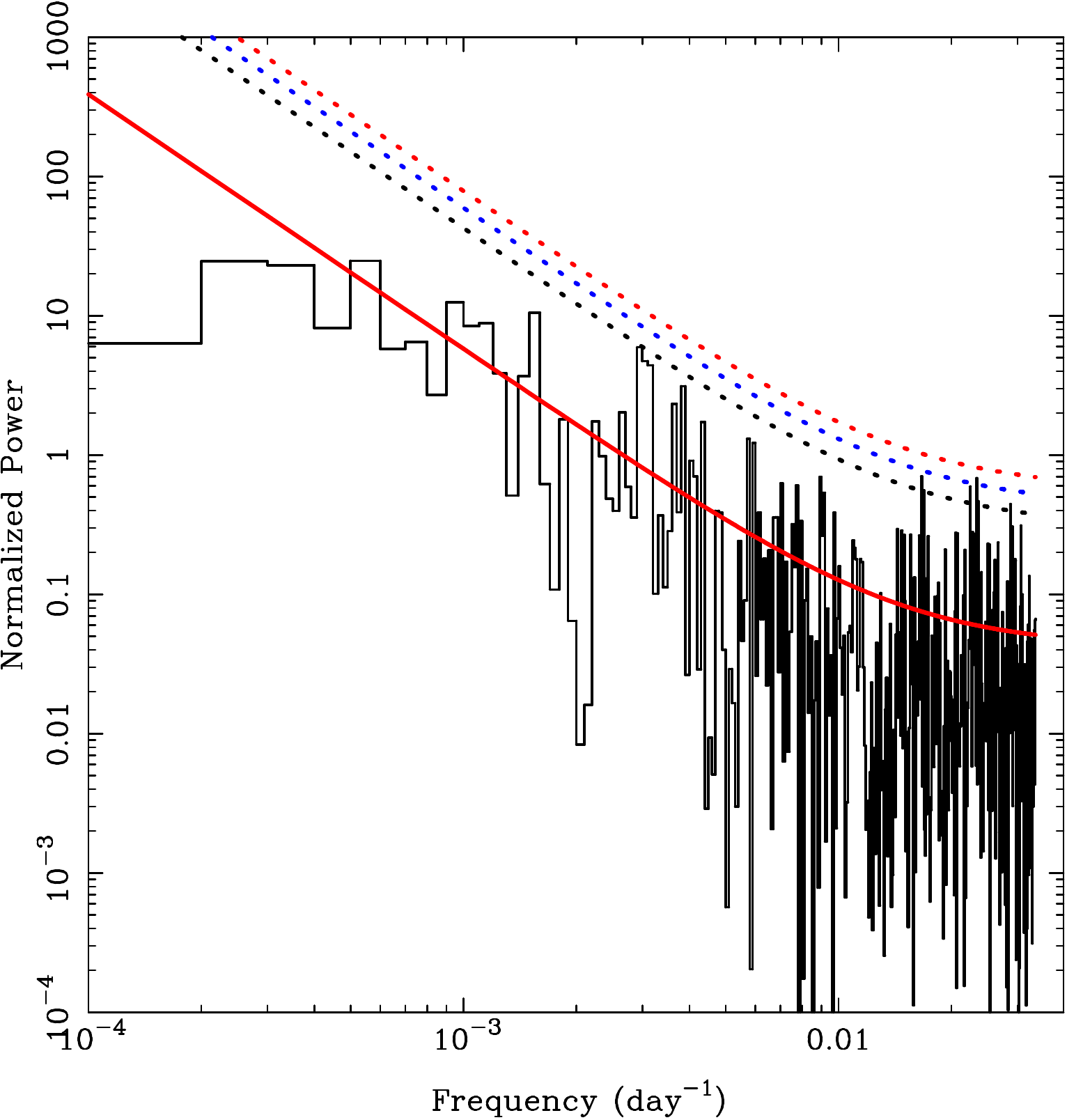}\hspace{1cm}\includegraphics[scale=0.5]{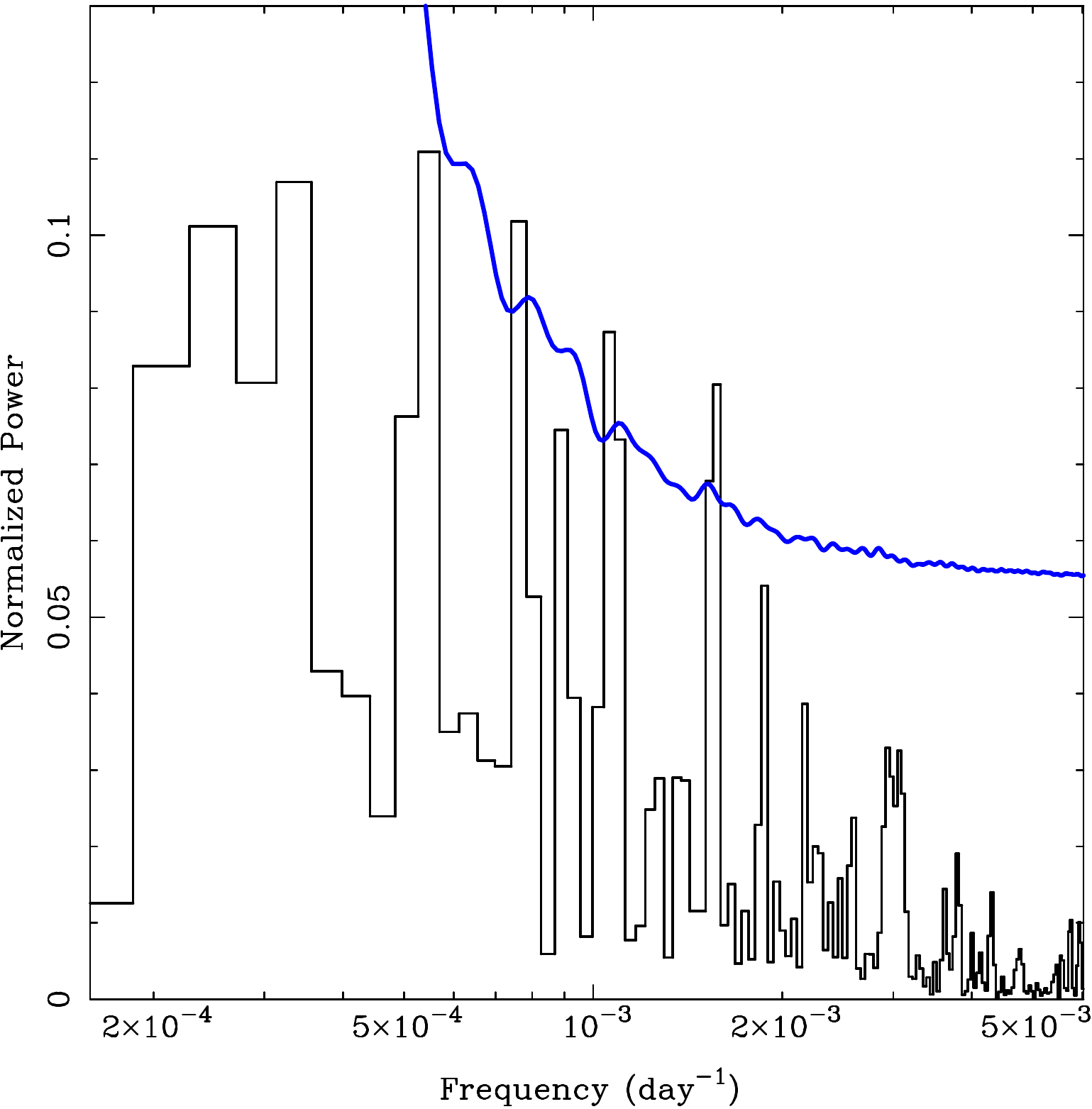}
\includegraphics[angle=90, scale=0.63]{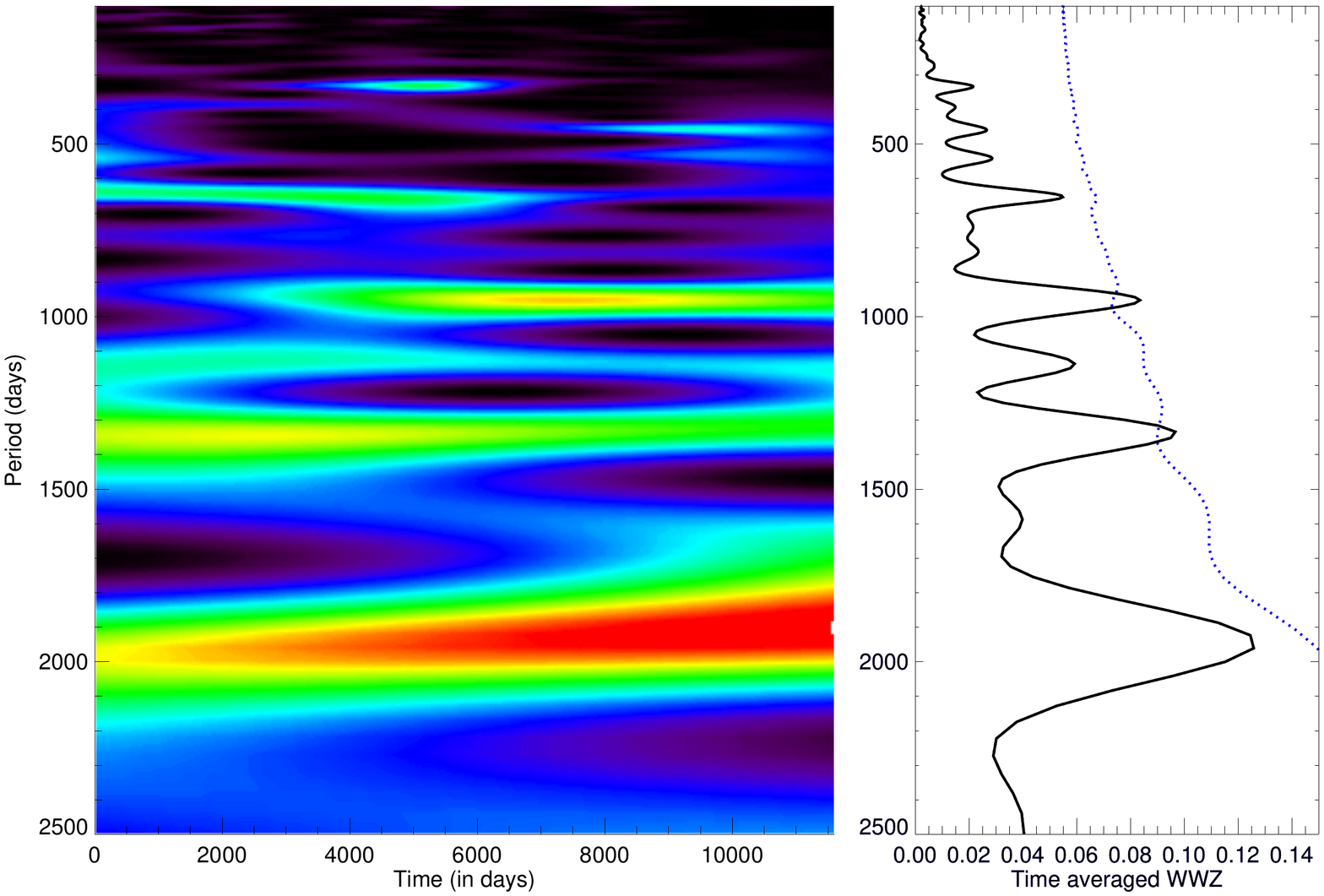}
\caption  {Time series periodicity analysis of AO 0235+164 for 4.8 GHz radio light curve. 
{\it Top left}: DCDFT analysis: the power spectral density curve is given in black and the best fit 
red noise spectrum is plotted in red. 
The black, blue and red dotted lines corresponds to $2\sigma$, $3\sigma$ and $4\sigma$ significance curves.
{\it Top right}: GLSP analysis: the normalized Lomb-Scargle power curve is given in black.  
The blue curve curve corresponds to $3\sigma$ significance.
{\it Bottom left}: WWZ power density plot (with red most intense and power decreasing toward violet and black).
{\it Bottom right}: Time averaged WWZ (in black); 99.73\% confidence levels are represented by
the dotted blue curve.}
\end{figure*}

\begin{figure*}
\centering
\includegraphics[scale=0.5]{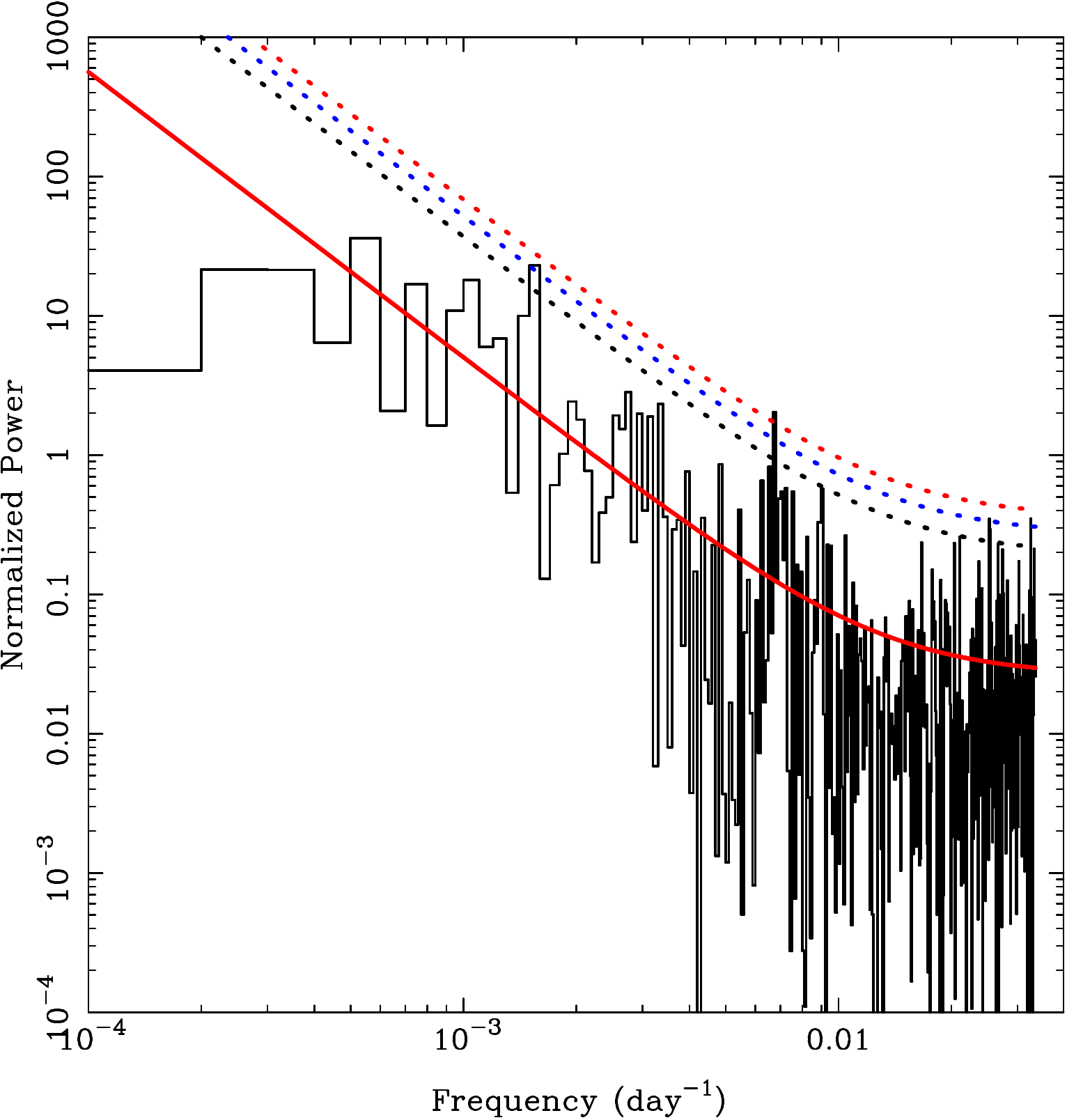}\hspace{1cm}\includegraphics[scale=0.5]{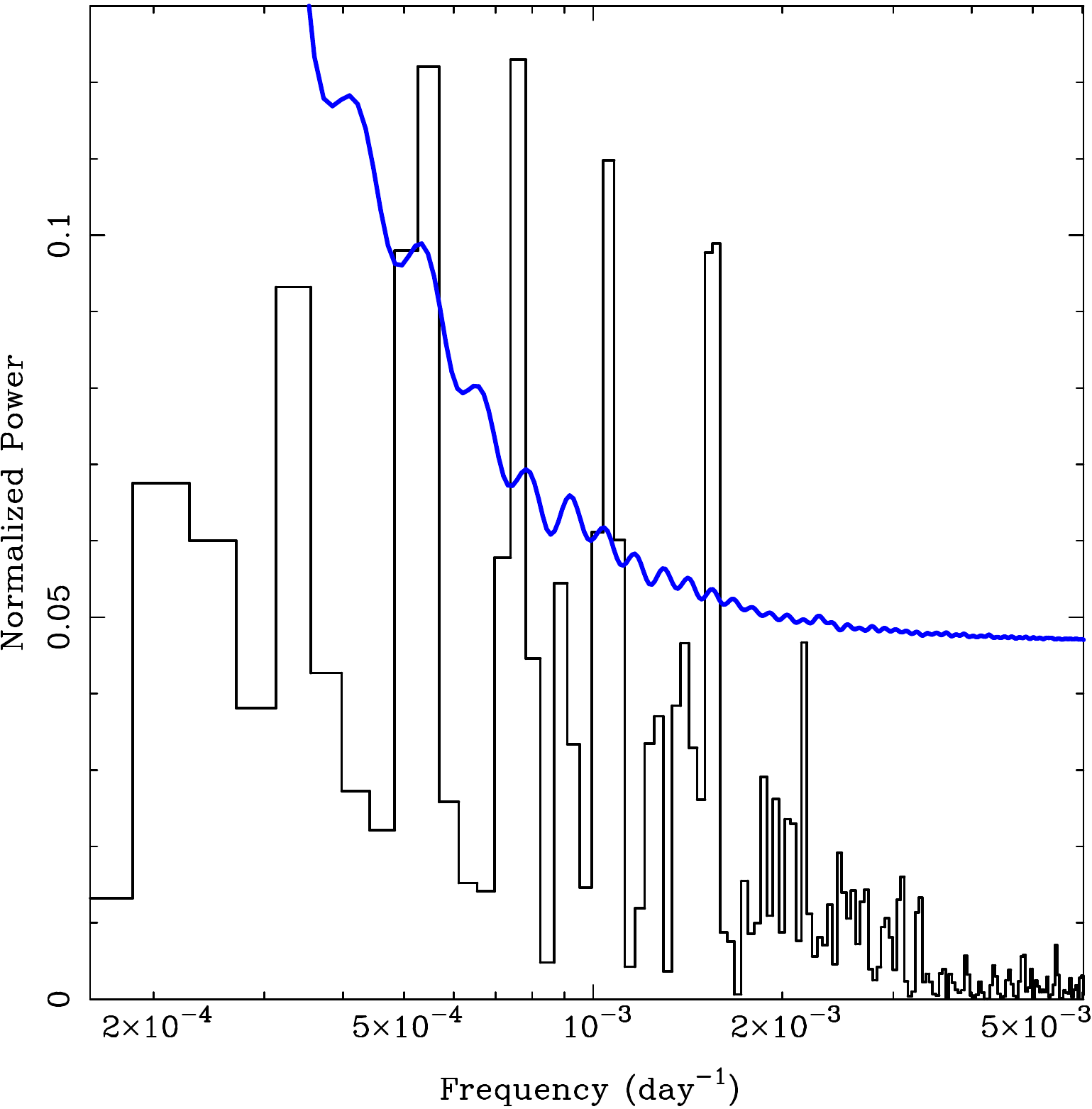}
%\vspace{-1cm}
\includegraphics[angle=90, scale=0.63]{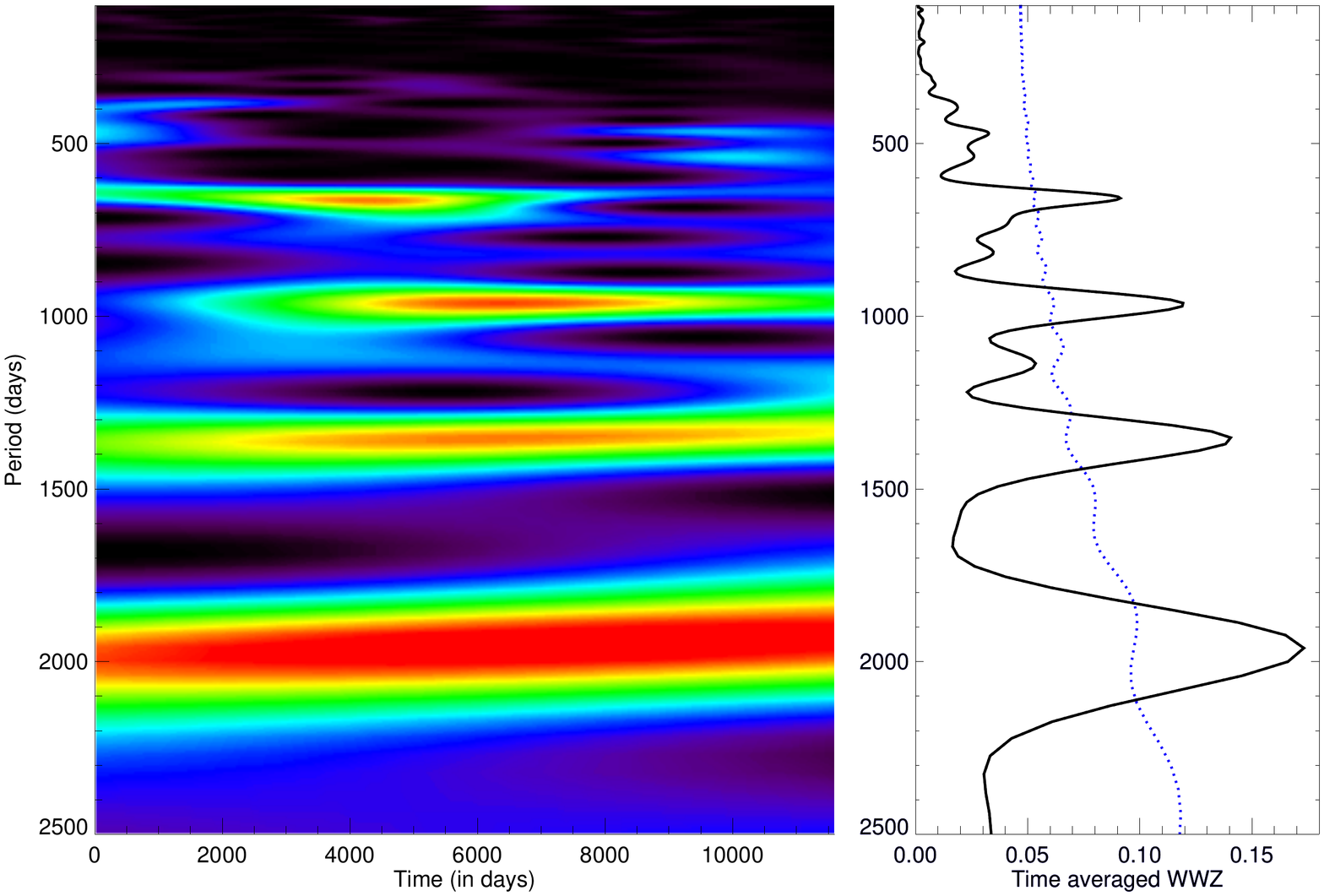}
\caption  {As in Fig.~2 for the time series periodicity analysis of AO 0235+164 for the 8.0 GHz radio light curve.}
\end{figure*}

\subsection{Data Compensated Discrete Fourier Transform}
\noindent
The estimation of the power spectrum for an unevenly sampled time series is one of the 
most common problems in searching for periodicities \citep[e.g.,][]{2017ApJ...837...45F}. The usage of the standard discrete Fourier
transform method for unevenly sampled data leads to complexities such as frequency shifting 
and amplitude fluctuations. Such problems can be ameliorated by using the DCDFT\footnote{The script is available at https://github.com/ilmimris/dcdft}. 
It differs from the usual Discrete Fourier Transform (DFT) in the model used to fit the 
light curve \citep{1995AJ....109.1889F}. Whereas the DFT uses sinusoidal components, the DCDFT uses the least square regression of sinusoidal and 
constant components. The effect of using only sinusoidal components (usual DFT) for unevenly spaced data could result 
in the discrepancy of 5\% for lower frequencies ($<0.02$ d$^{-1}$), at least in the case of a different observation \citep{1981AJ.....86..619F}. With a putative  QPO over
600 days ($\approx 0.0015$ d$^{-1}$), this difference might be significant.  The frequency spacing 
used by the DCDFT method is uniform for all three light curves regardless of gaps. \\ 
\\
At lower frequencies, 
the spectrum of blazars is, in general, dominated by red noise of the form $P\propto f^{-\alpha}$ at a frequency $f$, 
where $P$ is the PSD and 
$\alpha$, the spectral index. At higher frequencies ($>0.01$ Hz), the spectrum flattens and is dominated by white noise
($P\approx constant$). The red noise dominated component is fitted with a power-law plus constant using the whittle
likelihood function \citep{1953ArM.....2..423W, 2010MNRAS.402..307V}. The fit statistic for 
a given parametric model $P(f;\theta)$ for an observational dataset having
$N$ data points is given by twice the negative of the whittle likelihood function and is written as 
\begin{equation}
S = 2\sum_{j=1}^{N-2}\{\frac{I_j}{P_j}  + logP_j\},
\end{equation}
where $I_j$ is the periodogram at the frequency $j$. The best fit parameter $\theta$ is obtained by minimizing the 
whittle fit statistics.\\
\\
Here, we fit the power spectrum with the simple power law:
\begin{equation}
P(f) = Nf^{-\alpha} + C,
\end{equation}
where $N, \alpha$ and $C$ are normalization, spectral index, and constant, respectively.  Table 1 shows the best fit values of these parameters for all three light curves. We also tried fits with a bending power law in which there are additional parameters: a lower index and the
bend frequency $f_b$, the frequency where the spectral index of the power spectrum changes. The lowest frequency explored with the current data sets is
 is $\sim 8\times 10^{-5}$ d$^{-1}$.  The apparent bends in these power-spectra are $f_b < 4\times 10^{-4}$ d$^{-1}$ and well below any putative QPO discussed below. In addition, there is a strong degeneracy between any bend frequency and the normalization. So a simple power law is an adequate description for our purposes.\\

\begin{table*}
\centering
\begin{tabular}{|c|c|c|c|}
\hline
Light Curve & $\alpha$ & log$N$ & $C$\\ \hline
4.8 Hz & $1.83\pm 0.01$ & $-4.73 \pm 0.04$&$0.043\pm 0.001$\\
8.0 Hz & $2.09\pm 0.09$ & $-5.43 \pm 0.11$&$0.026\pm 0.002$\\
14.5 Hz & $1.84\pm 0.01$ & $-4.76 \pm 0.06$&$0.004\pm 0.001$\\     
 \hline\hline    
\end{tabular}
\caption{Best fit power-law values using the DCDFT method for the three radio datasets; $\alpha$ is the 
spectral index of the power-law, $N$ is the normalization, and $C$ is a constant.}
\end{table*}

The significance of any peaks are calculated using the method given in \citet{2005A&A...431..391V}. We can estimate the null
hypothesis and calculate the probability that any peak is produced by a non-periodic noise signal.
The estimation of null hypothesis gives the false alarm probability (FAP) represented by a
$p$ value.
The significance level $Y$ at a given FAP $p$ for $n$ number of frequencies considered is calculated as 
\begin{equation}
Y = -ln[1 - {\rm pow}((1-p,1/n'))],
\end{equation}
where $n'= n-2$ \cite{2005A&A...431..391V}. The confidence factor (C.F.) can be estimated from the value of $p$ as
\begin{equation}
C.F. = (1-p)\times 100.
\end{equation}
The C.F. can be interpreted in terms of standard deviations ($\sigma$) using the cumulative distribution 
function.\\
\\
The significance for a DCDFT is calculated with respect to the 
red noise spectrum, so the significance curves have the same slope, $\alpha$, as the PSD.

\subsection{Generalized Lomb-Scargle Periodogram}

We use the Generalized LSP 
(GLSP) routine of the {\sc pyastronomy} package\footnote{https://github.com/sczesla/PyAstronomy} 
which implements the algorithm mentioned in \citet{2009A&A...496..577Z}. In addition to the `classical' 
Lomb-Scargle algorithm (\cite{1976Ap&SS..39..447L, 1982ApJ...263..835S}), it takes into account 
the errors associated with the fluxes. It also uses sinusoids plus constant as a fitting function unlike the classical LSP where only 
sinusoidal functions are employed. Frequency binning depends on the oversampling factor. Here, 
we took the number of frequencies to be half of the number of data points in a particular observation. \\
\\
As we discussed before, the power spectral density of a blazar is usually well described by a negative spectral index.
This implies that the spectrum oscillates with large amplitude at 
low frequencies which could be mistaken as QPO peaks. Besides, the noisy nature of a periodogram can also 
be misinterpreted as possessing peaks and requires a careful analysis. In order to deal with these issues, the significance
of potential QPO peaks can be estimated by simulating light curves of a given power spectrum. \\
\\
For these simulations, we
assumed that blazar light curves can be fully characterized by the overall stochastic processes occurring in the accretion disc 
and turbulent jets. This assumption is supported by the observations that the PSD of multi-frequency blazar light curves
can be best represented by power-law shape (e.g., \cite{2012arXiv1205.0276M}, radio; \cite{Nilsson:2018tet}, optical, and 
\cite{2020ApJ...891..120B}, gamma-ray). But the process giving rise  to any QPOs, which most likely arise from a coherent 
disc or jet processes \citep[see][]{2015ApJ...813L..41A}, can be distinguished from such overall stochastic variability in the sense
 that the features due to QPO should stand out above the power-law shape of the PSD. This idea forms the basis for the estimation
  of the significance of the observed QPOs and thereby directly affects our conclusions. However, it should be recalled that at 
  the lowest frequencies probed by a given dataset, where the variability due to stochastic processes is larger, it is more 
  difficult to disentangle the PSD signatures due to putative QPOs (see \cite{2020ApJ...891..120B} for discussion).\\
\\
The simulations are done by randomizing the amplitude and phase of the Fourier components of the light
curve with the given source properties; for details, see \cite{1995aap}. We used the same time sampling, number of data 
points and mean flux density as the original light curves for these simulations. For the power spectrum, we use the best-fit 
values and the associated uncertainties obtained in the DCDFT fitting procedure for a simple power-law model using Whittle 
likelihood function.
We use the {\sc stingray} package\footnote{https://github.com/StingraySoftware} which implements the approach of \cite{1995aap} 
to simulate 1000 light curves with same properties as the source light curve and of a given power spectrum. 
The software takes into account the uncertainties of the parameters
  and thus allows for a range of parameters during the simulations. \\
\\
We calculate the GLSP
for the simulated light curves. In order to take the effect of frequency sampling in account, 
we use the mean of these simulated LSP at each frequency as the baseline and estimate the $3\sigma$ significance using Eqns.\ (3) and (4). Only employing the spectral distribution of an LSP at each frequency for calculating significance does not
properly include the number of frequencies examined and thus is expected to overestimate the significance.

\subsection{Weighted Wavelet Z-transform  Analysis}

\noindent
It is possible for a QPO signal to not be persistent throughout the data train because the signals evolved both in
amplitude and frequency with time. In order to assess the non-stationarity of the signal, it should best be  decomposed 
into frequency/time--space simultaneously \citep{1998Bulletin}. For this purpose, the wavelet technique is widely 
used: the data are compared to wavelet functions instead of sinusoids.
Given the uneven sampling nature of the data used, an improved form of the wavelet transform is best used. 
The weighted wavelet Z-transform (WWZ) is a widely used wavelet technique which employs $z$-statistics 
\citep{1996AJ....112.1709F}.  It enhances the performance of the wavelet technique for the 
realistic data which could be sparse and unevenly sampled \citep{2005NPGeo..12..345W}. It is based on the Morlet wavelet function which
varies with both frequency and time \citep{1989wtfm.conf....2G}. We use WWZ software\footnote{WWZ software is available at URL:  
https://www.aavso.org/software-directory.} to calculate the WWZ power for a given time and 
frequency as has often been done in the recent literature \citep[e.g.,][and references therein]{2013MNRAS.436L.114K,2016ApJ...832...47B,
2017ApJ...847....7B,2017aApJ...835..260Z,2017bApJ...842...10Z,2018ApJ...853..193Z}. \\
\\
The major underlying assumption in WWZ analysis is the choice of wavelet function. In addition, frequency
and space resolution also plays an important role. Here, we used the same frequency resolution for 
all three radio frequency light curves. Due to the finite length of the time series, 
edge effects could arise in the density plot which will be seen as spurious peaks. This
effect should also considered while calculating the WWZ power and significance of peaks.\\
\\
We also calculated the power spectral density of the WWZ which gives the same information as WWZ in frequency 
space, marginalizing over the length of the observation. Since it is a periodogram, the PSD would usually follow the power-law
and distributed as $\chi^2$ with 2 degrees of freedom \cite{2005A&A...431..391V}. 
For significance estimation, we calculated WWZ and time averaged WWZ for 1000 simulated light curves with properties the same 
 as source light curve, as done in \citet{2017ApJ...847....7B}. We used Eqns. (3) and (4) to estimate the 99.73\% confidence interval ($3\sigma$)
 using the mean of simulations as baseline and including the number of frequencies in the calculation of significance.

\begin{table*}
\centering
\begin{tabular}{|c|c|c|c|c|c|c|c|c|c|c|}
\hline
Light Curve & Method & 1st peak & Significance & 2nd peak  & Significance & 3rd peak & Significance&4th peak &Significance\\ \hline
\multirow{3}{*}{4.8 GHz} & GLSP & $644^{+17}_{-17}$ &$3\sigma$--$4\sigma$ & $962^{+39}_{-18}$& $\sim 4\sigma$ & $1349^{+93}_{-72}$& $3\sigma$--$4\sigma$
                         &$1898^{+66}_{-151}$&$2\sigma$--$3\sigma$\\
                        & DCDFT& $667^{+47}_{-42}$ & $<2\sigma$&-&-&-&-&-&-\\
                        & WWZ &  $654^{+4}_{-4}$   & $2\sigma$--$3\sigma$&$952^{+9}_{-9}$&$3.5\sigma$--$4\sigma$& $1333^{+33}_{-18}$& 
                        $3\sigma$--$4\sigma$ &$1961^{+39}_{-37}$&$2\sigma$--$3\sigma$ \\ \hline
\multirow{3}{*}{8.0 GHz} & GLSP & $653^{+16}_{-19}$ & $5\sigma$--$6\sigma$ & $948^{+31}_{-30}$& $5\sigma$--$6\sigma$ & $1337^{+64}_{-58}$& $\sim 5\sigma$
                         & $1964^{+141}_{-123}$& $\sim 5\sigma$\\    
                         & DCDFT& $667^{+47}_{-42}$ & $\sim 3\sigma$&-&-&-&-&-&-\\    
                         & WWZ &  $658^{+5}_{-4}$   & $\sim 5\sigma$ & $971^{+10}_{-9}$& $5\sigma$--$6\sigma$& $1351^{+12}_{-18}$& $5\sigma$--$6\sigma$
                         & $1961^{+39}_{-37}$& $ 5\sigma$--$6\sigma$ \\ \hline   
\multirow{3}{*}{14.5 GHz} & GLSP & $663^{+19}_{-18}$ & $5\sigma$--$6\sigma$& $963^{+39}_{-38}$ & $5\sigma$--$6\sigma$
                         & $1352^{+83}_{-73}$ & $\sim 5\sigma$ &$1899^{+167}_{-143}$&$\sim 5\sigma$\\    
                         & DCDFT& $667^{+47}_{-42}$ & $\sim 3\sigma$&-&-&-&-\\    
                         & WWZ &  $658^{+5}_{-4}$   & $4\sigma$--$5\sigma$ & $971^{+10}_{-9}$&$5\sigma$--$6\sigma$ & $1370^{+10}_{-19}$&$\sim 5\sigma$
                         & $1961^{+39}_{-37}$& $5\sigma$--$6\sigma$ \\ \hline               
 \hline\hline    
\end{tabular}
\caption{Results of the periodicity analysis of AO 0235+164 UMRAO radio
observations centered at 4.8, 8.0 and 14.5 GHz.  The quoted significances are based upon comparisons of the peaks with 3-, 4-, and 5-$\sigma$ curves. The n$\sigma$--(n+1)$\sigma$ type of entry in the table means that the significance clearly exceeds n$\sigma$ but does not reach (n+1)$\sigma$.}
\end{table*}

\begin{figure*}[t]
\centering
\includegraphics[scale=0.5]{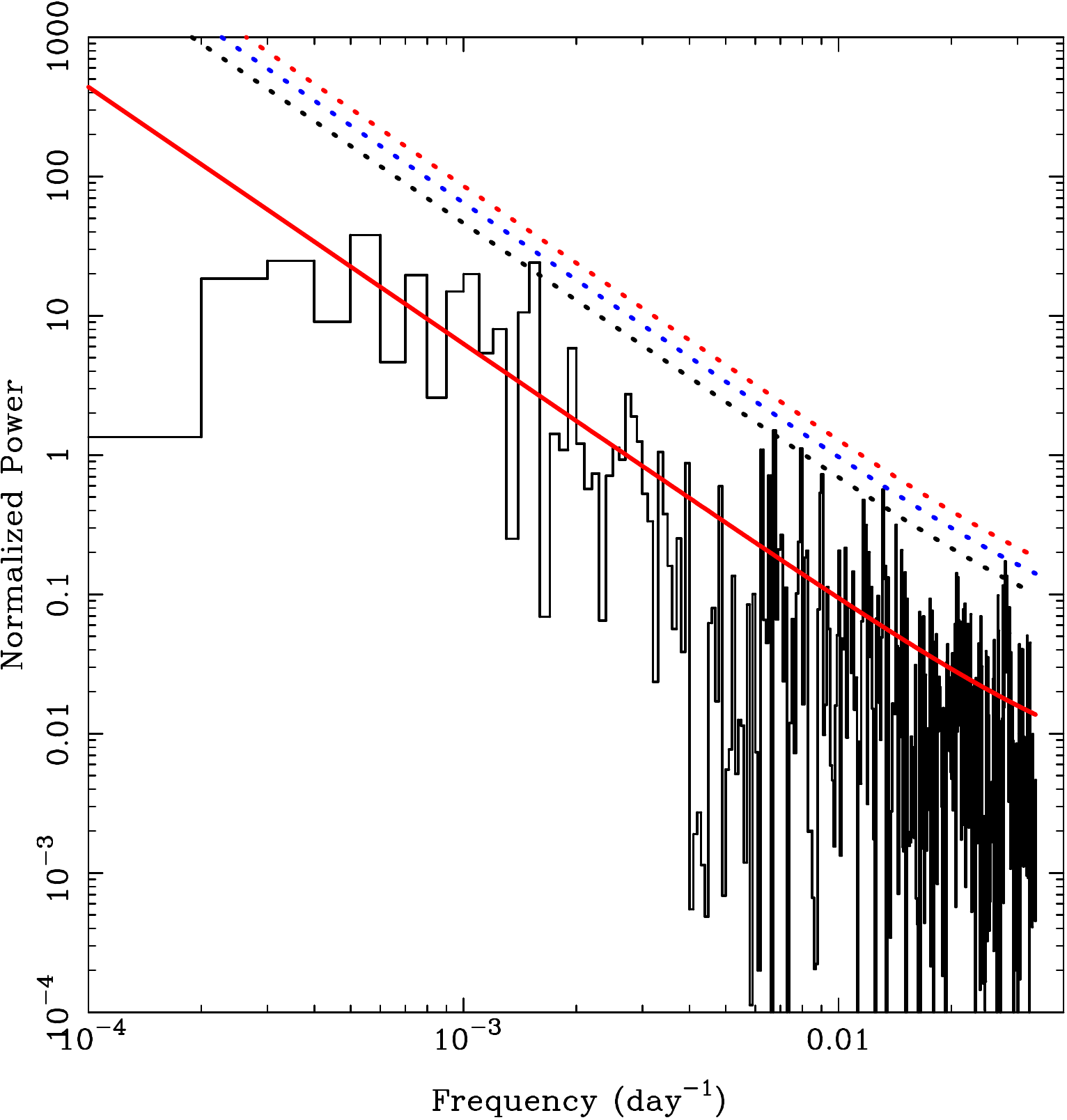}\hspace{1cm}\includegraphics[scale=0.5]{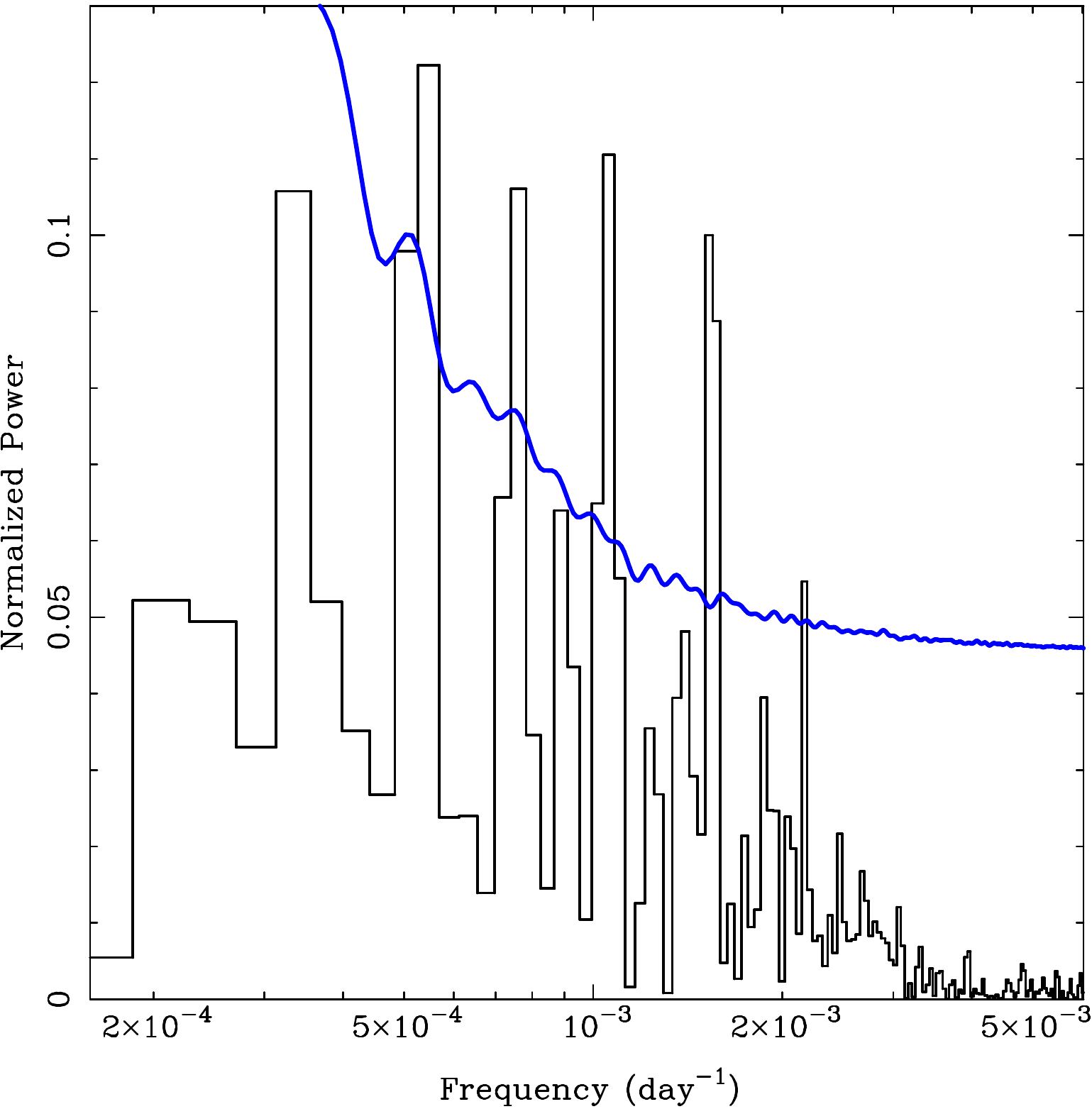}
%\vspace{-1cm}
\includegraphics[angle=90, scale=0.63]{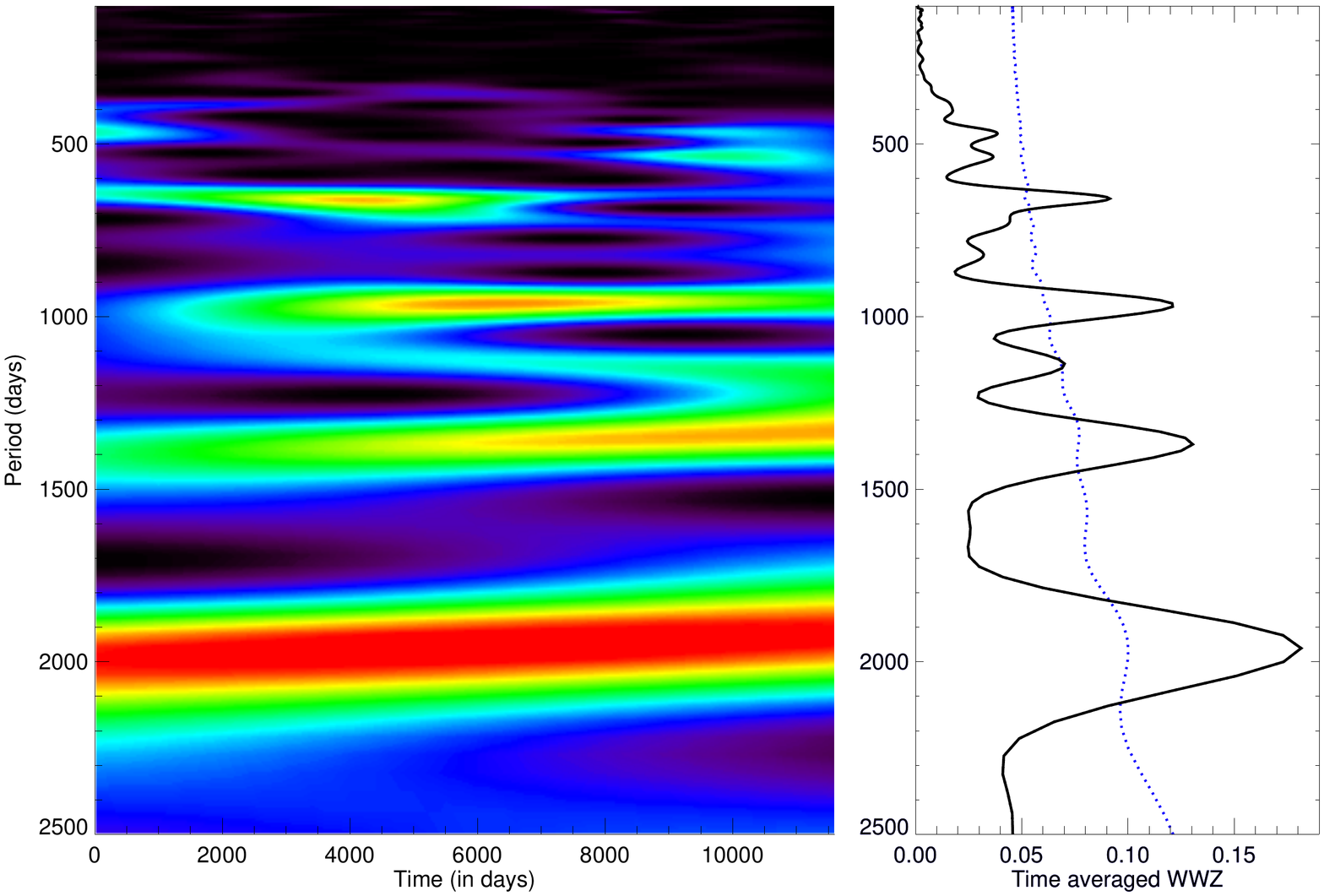}
\caption  {As in Fig.~2 for the time series periodicity analysis of AO 0235+164 for the 14.5 GHz radio light curve.}
\end{figure*}

\section{Results}
\noindent
Fig.\ 2 shows the DCDFT, GLSP and WWZ analyses of the 4.8 GHz radio 
light curve shown in the upper panel of Fig.\ 1. 
The upper left panel shows the DCDFT power
plotted against frequency. The red curve denotes the best-fit red noise spectrum. 
The black, blue and red dotted curves correspond to $2\sigma$, $3\sigma$ and $4\sigma$ levels of 
significance respectively.
In these plots, the effect of the white noise has been significantly reduced by using the DCDFT, which employs 
sinusoids plus constants in the fit. For this 4.8 GHz DCDFT plot, the peak around $667 \pm 50$ days shows less than  
$2\sigma$ significance.  Another peak around 342 days seems to be more significant but still only about $2\sigma$. 
The central values, spreads, and significances of the various peaks are given in Table 2.\\
\\
The upper right panel of Fig.\ 2 shows the GLSP power plotted against frequency for the 4.8 GHz observations (black curve). The blue
curve corresponds to the $3\sigma$ significance calculated using simulations. There are three peaks 
at around 644, 962 and 1348 days whose significance is more than $3\sigma$. As compared to the DCDFT analysis,
the significance of the peak around 344 days decreases and that around 644 days increases. However, the significance is highest for a nominal peak period around 962 days. \\
\\
In order to confirm the significance of any putative QPO, we need to examine its persistent throughout these lengthy observations.
For this purpose, the WWZ approach is critical as 
it can quantify how long a signal is present in the observation. The lower left panel of Fig.\ 2 shows the
WWZ density plot in the time-frequency plane. 
In this density plot, we found considerable power concentrations around 953, 1334 and 1960 days persisting  
throughout the observations.
A feature around 654 d is present 
but its significance is less than $3\sigma$ and it is present for lesser duration
as compared to other three features. The lower right panel shows the time averaged WWZ plotted 
against the period to estimate the significance of the periodic feature. The blue dotted line represents
99.73\% significance. We found the three peaks centred at around 952, 1333 and 1961 days having more than $3\sigma$ significance,
with the first being more significant. The peak around 342 d is not significant at all.\\   
\\
The features seen in the WWZ at $\sim 954$ and $\sim 1961$ days are more than $3\sigma$ significant and also persist the longest at high strength throughout
the observations. 
As the WWZ power takes into account the strength of the signal in both frequency and time, the presence of 
the $\sim 1961$ d feature for longer lengths of time makes it appear most significant at 4.8 GHz, but it exhibits fewer cycles and in that sense may be less convincing and so it may be better considered as a 1:2 subharmonic of the $\sim 952$d signal.  The feature around  654 days could be in the ratio of 3:2 of that one, as has been seen in 
many XRB sources \citep{2018A&A...616L...6G}. 
We note that a $\sim 25$ year long set of this UMRAO data (starting 5 years earlier and ending 10 years earlier) gave some indication of a $\sim$ 3.7 y QPO \citep{2003APSC..300..159A}, which is consistent with the 1333 d signal, or about twice the one around $654$ d.  Other previous claims of possible QPOs in AO 0235+164 are mentioned in Section 5. \\
\\ 
Fig.\ 3 shows the time series analyses of the 8.0 GHz radio light curve with 
DCDFT, LSP and WWZ. The DCDFT shows the feature at $667 \pm 50$ 
days with a $3\sigma$ detection and shows no other peak with such high significance. However, the GLSP method displays four features centred at 653, 948, 1337 and 1964 days having significance 
more than $4\sigma$, with the first peak being slightly stronger.  The WWZ analysis confirms these quasi-periodicities indicated by GLSP, though with slightly different nominal peaks, given in Table 2.
The peak around $658\pm 5$ days is detected with more than $4\sigma$ significance. However, and similarly to the 
4.8 GHz light curve, even stronger signals are detected centred at 971, 1351 and 1961 days. In this case,
the two  longer periods are found to be more significant than the lower ones, with the peak around 1351 days being the strongest. \\
\\
The results for the 14.5 GHz light curve shown in the lowest panel of Fig.\ 1 
are presented in Fig.\ 4. Similarly to the 4.8 GHz and 8.0 GHz light curves analysis, the DCDFT method gives 
the most significant peak ($\sim 3\sigma$ significance) at $667 \pm 50$ d. As for the 8.0 GHz data, no comparably 
significant peak is detected. The GLSP analysis yields similar peaks to those at the other bands, with significances of more than $3\sigma$ centred at 663, 963, 1352 and 1899 days.  The signal at lower peaks (at 663 and 963 days) are found to be stronger, with 963 as the strongest. 
In addition to these, a less significant feature at 465 days is also detected.  
The WWZ analysis supports the results for the other bands, and the peak with more than $4\sigma$ confidence
 is been detected at $658\pm 4$ d with a comparable significant signal at 971, 1370 and 1961 d. Among all peaks at 14.5 GHz, 
 the strongest signal is detected at the period of 971 days. \\
 \\
 The harmonics at lower frequencies ($\sim 10^{-3}$ d$^{-1}$) are not displayed 
 in the DCDFT analysis as the frequency 
 resolution for that approach is very low there. The fact that this peak around 660 days is detected by all methods,
 supports the hypothesis that the other peaks detected with less or more significance could be the harmonics of the 
 this period. However, for the 4.8 GHz light curve, the next longer period, of $\sim 965$ d is the most significant in both the GLSP and 
 WWZ analysis. In addition, for the other two light curves, this period is consistently very highly ($\sim 5.5\sigma$) significant by both 
 methods. 
Unsurprisingly for these data sets, where all three wavelengths are sampled at the same intervals and for the same span, the putative QPO periods from a particular method are essentially the same for all three radio light curves because of the same 
frequency resolution.

\section{Discussion and Conclusions}
\noindent
In this paper, we have analyzed over three decade-long total flux density observations of the blazar AO 0235+164 made at the primary radio
frequencies of 4.8, 8.0 and 14.5 GHz at UMRAO.
From visual inspection, there seemed to be a distinct possibility of detecting  periodicities in the light curves. 
For the analysis, we used observations taken in all three bands over the same time interval (1980 through 2012) as shown in Fig.\ 1. We used methods involving different approaches 
to  measure the periodicities in the light curves. 
In order to produce a robust calculation of the significance of the periodic peaks, one needs to model the lower-frequency portion of the spectrum 
as a red-noise which means that the power density is inversely proportional to a power of the frequency. Assuming the
spectrum consists of red noise and white noise at higher frequencies, the observed spectra is fitted with these two components. 
Moreover, global significances are calculated as they give the significance with respect to the whole period of the observations. \\
\\
Combining the results from the different approaches, a quasi-period of $\sim 665$ days is detected with at least $3\sigma$ significance with the  GLSP, DCDFT and WWZ approaches for the 8.0 and 14.5 GHz radio light curves.
For 4.8 GHz, the same peak is detected by DCDFT but with lesser significance. 
With the WWZ and GLSP analyses of all three light curves, we also find detections of periodicities 
around 965, 1350 and 1960 days.  In these, somewhat more robust, approaches these longer periods all have comparable or more significant signals than the 665 day one, particularly the $\sim 965$ d (or 2.64 y) peak, and we tentatively conclude that this one is the fundamental mode and note the 667 d signal is very close to a 3:2 frequency harmonic.  Although the plots of the WWZ results in the lower-left panels of Figs.\ 2--4 show that the $\sim$1960 d peak is  stronger than the one at $\sim$965 d, because it covers fewer than 6 cycles even in these very lengthy measurements, it is not quite as significant when integrated over time, as indicated in the lower-right panels, and as listed in Table 2. \\
\\
There have been earlier QPO claims for AO 0235+164 in optical and radio
bands. Periodicities of 1.29, 1.53 and 2.79 y have been suggested to be present in early optical
light curves of this source \citep{1988AJ.....95..374W}. Later, periods of 2.7 and 3.6 y
were also suggested \citep{1995PASP..107..863S} as 
were claims of events with periodicity of $\sim 11.2$y in optical wavebands that might reflect an underlying $\sim 5.7$y QPO \citep{2001A&A...377..396R}. 
\citet{2002A&A...381....1F, 2007A&A...462..547F}
claimed the detection of optical QPOs with periodicities of 2.95 years and 5.87 years. 
In the radio band, a  quasi-period of $\sim 5.7$ y  was reported for the 8.0 GHz data \citep{2001A&A...377..396R} but was based on much more limited data; it might be
 consistent with twice the strongest period claimed in this work.
\citet{2002A&A...381....1F, 2007A&A...462..547F} suggested a period of 10 years at 4.8 GHz, 5.7 years at 8.0 GHz
and 5.8 years at 14.5 GHz, again based on shorter light curves than considered here.\\ 
\\
As the total emission in radio loud AGN is comprised 
of contributions from the accretion disc and jet, QPOs in some bands can, in principle, arise from either of those regions.
But AO 0235+164 is historically classified as BL Lac type of blazar, and in such objects the non-thermal emission dominates the total
emission across the electromagnetic spectrum \citep[e.g.,][]{2019MNRAS.484.5785G}, and regardless of any disc contribution in some bands the radio emission we are
investigating here certainly comes from the jet.  Even at other wavelengths, because the jet is viewed  at a very small inclination angle 
\citep{1995PASP..107..803U} the overall synchrotron emission from the jet is amplified by the substantial relativistic beaming \citep[e.g.,][]{2001ApJS..134..181J}, thereby
overwhelming any quasi-thermal emissions coming directly from the accretion disc which are not so beamed.  
Jet variability in general is enhanced by strongly relativistic motions often revealed by apparent superluminal motions of the knots
where the Doppler boosting factor depends sensitively on both the flow speed down the jet
and the precise viewing angle toward it, so modest inherent changes in either can be observed as major flux changes.  In addition, the observed
temporal variations are compressed by that Doppler factor from those emitted in the jet frame. \\  
\\ 
One likely route to QPOs arising in the jets of blazars 
 involves an  internal helical structure \citep[e.g.,][]{1992A&A...255...59C, 1999A&A...347...30V, 
2004ApJ...615L...5R, 2015ApJ...805...91M}.  Such structures can be  inferred from the wiggling of knots observed in VLBI jets that can
be resolved transversely at different epochs \citep[e.g.][]{2010ApJ...723.1150P, 2020A&A...634A..87L}.  They can be understood in terms of shocks
propagating outward through the jets and interacting with a preexisting  helical structure, presumably related to the toroidal
component of the jet magnetic field, even in an otherwise cylindrical or conical jet.
This magnetization of the jets can be related to the winding up of the  
magnetic field around the rotating SMBH which would have its origin in accretion flow with accumulated magnetization, although at the parsec scales where the radio emission arises there
is almost certainly an important turbulent contribution to the magnetic field.  Instabilities leading
to sudden changes in the magnetic field strength and/or matter density can lead to the formation of periodic oscillations at the magnetosphere above the accretion disk
that might be advected into the jet base 
\citep{1992A&A...255...59C, 2004ApJ...601..414L, 2012MNRAS.423..831F}.  Magnetohydrodynamic simulations of jets support the likelihood of magnetic fields
playing a major role in the origin of these QPOs
\citep[e.g.][]{2012MNRAS.423.3083M}, including the recently suggested possibility of dominant kink-instabilities in the jet spine \citep{2020MNRAS.494.1817D}. \\     
\\
Alternatively, a modest amount of bulk jet precession could naturally produce an observed QPO through the resulting changes to the Doppler boosting. This wiggling could be 
intrinsic to the source \citep[e.g.][]{1992A&A...259..109G},   where it might be produced by Lense-Thirring precession of the inner portions of the accretion disk produced from frame-dragging around a rotating SMBH \citep{1998ApJ...492L..59S, 2018MNRAS.474L..81L}. Precession can also naturally be produced by 
a binary SMBH companion \citep[e.g.,][]{2017A&A...602A...29B, 2018MNRAS.478..359K} and would be observed as wiggling of the jet that would provide a quasi-periodic contribution to the
variability.  Purely hydrodynamical Kelvin-Helmholtz instabilities can generate weak shocks via growing and then damping of natural oscillations in the flow if there is some amount of jet precession \citep{2001ApJ...555..744H}, and these can yield  QPOs \citep{2003APSC..300..159A}.
Either the helical structure within a jet or the orbital-induced precession of a jet appear to be viable explanations for these QPOs of $\sim 960$d  and $\sim 660$d and that seem to be present in the radio emission of AO 0235+164, but a Lense-Thirring precession origin is unlikely to produce a period longer than a few months \citep{1998ApJ...492L..59S}. 

\section*{Acknowledgements}
\noindent
We would like to thank the referee for useful comments and suggestions to improve the manuscript. ACG is thankful to Gopal Bhatta for discussions. The work of AT and CB was supported by the Innovation Program of the Shanghai Municipal Education Commission, Grant No.~2019-01-07-00-07-E00035, and the National Natural Science Foundation of China (NSFC), Grant No.~11973019. ACG is partially supported by Chinese Academy of Sciences (CAS) President's International Fellowship Initiative (PIFI) (grant no. 2016VMB073).
 MFG is supported by the National Science Foundation of China (grant 11873073).  
 This research is based on data from the University of Michigan Radio Astronomy Observatory, which was supported by
the National Science Foundation, NASA and the University of Michigan.  UMRAO research  was supported in part by a series of grants from the NSF (most recently
AST-0607523) and by a series of grants from NASA including Fermi G.I.\ awards NNX09AU16G, NNX10AP16G, NNX11AO13G, and NNX13AP18G.

\section*{Data Availability}

The total flux density data used in the paper will be available on reasonable request to Ashutosh Tripathi (Email: ashutosh\_tripathi@fudan.edu.cn).

\end{document}